                \documentclass[preprint2]{aastex}

\slugcomment{to appear in ApJ }

\shorttitle{Spectrum of Galactic Cosmic Rays Accelerated in SNRs}

\shortauthors{Ptuskin et al.}

\begin{document}

\title{Spectrum of Galactic Cosmic Rays Accelerated in Supernova Remnants}

\author{Vladimir Ptuskin and Vladimir Zirakashvili}

\affil{Pushkov Institute of Terrestrial Magnetism, Ionosphere
 and Radio Wave Propagation of the Russian Academy
 of Science (IZMIRAN), Troitsk, Moscow Region 142190, Russia}

\and

\author{Eun-Suk Seo}

\affil{Department of Physics and Institute of Physical Science and
Technology, University of Maryland, College Park, MD 20742 USA}

\begin{abstract}
The spectra of high-energy protons and nuclei accelerated by
supernova remnant shocks are calculated taking into account
magnetic field amplification and Alfvenic drift both upstream and
downstream of the shock for different types of supernova remnants
during their evolution. The maximum energy of accelerated
particles may reach $5\cdot10^{18}$ eV for Fe ions in Type IIb
SNRs. The calculated energy spectrum of cosmic rays after
propagation through the Galaxy is in good agreement with the
spectrum measured at the Earth.

\end{abstract}

\keywords{acceleration of particles --- shock waves --- supernova
remnants}

\section{Introduction}

Supernova remnants (SNRs) are recognized as the principal sources
of Galactic cosmic rays, and the diffusive shock acceleration is
accepted as the mechanism of cosmic ray acceleration by a
supernova blast wave moving through the turbulent interstellar
medium. Accelerated by supernova shocks, the energetic particles
diffuse in the interstellar magnetic fields and fill the entire
Galaxy. Clear evidence for particle acceleration in SNRs is given
by observations of non{}-thermal radio, X{}-ray, and gamma{}-ray
radiation. It is experimentally established from H.E.S.S. (High
Energy Stereoscopic System) data that there are cosmic{}-ray
particles with energies exceeding $10^{14}$ eV in the shell of the
supernova remnant RX J1713.7{}-3946 \citep{Ahar07}.

For a time, the theoretical estimates of maximum proton energy
were at a level of  $E_{\mathrm{max}}{\approx}10^{13}-10^{14}$ eV
\citep{Lag83}. The work of \citet{Bell04} demonstrated that a very
large random magnetic field $\mathrm{{\delta}B}\gg
B_{\mathrm{ism}}$ (where $B_{\mathrm{ism}} {\approx}
5\mathit{{\mu}G}$ is the average interstellar field upstream of
the shock) can be generated by a cosmic{}-ray streaming
instability in the precursor of strong shocks. It results then in
efficient confinement of energetic particles in the shock
vicinity, and it may raise the maximum particle energy by about
two orders of magnitude (the exact value depends on the supernova
parameters, see below). Remarkably, data on synchrotron X{}-ray
emission from a number of young SNRs proved the presence of strong
magnetic fields  $150$  to $500$ ${\mu}$G \citep{Voelk05} that can
be naturally explained by the effect of cosmic{}-ray streaming
instability. An excellent concise review on the origin of observed
cosmic{}-ray spectrum was presented by \citet{Hillas06}.

In the present work we calculate the steady state spectrum of
cosmic rays in the Galaxy using recent results on magnetic field
amplification in SNRs and including different types of SNRs in the
consideration. A numerical code is employed in the simulations of
particle acceleration and the SNR shock evolution.

\section{Modelling of cosmic ray acceleration in supernova remnants}

Because of high efficiency of shock acceleration, the spectrum of
cosmic rays should be selfconsistently determined with the account
of shock modification caused by the pressure of accelerated
particles. We are studing cosmic ray acceleration and the
evolution of a supernova blast wave with the use of our numerical
code described in \citep{Zir08a,Zir09}. The hydrodynamic equations
are solved together with the diffusion{}-convection transport
equation for the cosmic ray distribution function $f(t,r,p)$,
which depends on time $t$, radial distance from the point of
supernova explosion $r$ (here the spherical symmetry is assumed),
and the particle momentum $p$. The full system of equations is the
following:
\begin{equation}
\frac {\partial \rho }{\partial t}=-\frac {1}{r^2}\frac {\partial
}{\partial r}r^2u\rho,
\end{equation}

\begin{equation}
\frac {\partial u}{\partial t}=-u\frac {\partial u}{\partial
r}-\frac {1}{\rho } \left( \frac {\partial P_g}{\partial r}+\frac
{\partial P_c}{\partial r}\right),
\end{equation}

\begin{equation}
\frac {\partial P_g}{\partial t}=-u\frac {\partial P_g}{\partial
r} -\frac {\gamma _gP_g}{r^2}\frac {\partial r^2u}{\partial r}
-(\gamma _g-1)(w-u)\frac {\partial P_c}{\partial r},
\end{equation}

\[
\frac {\partial f}{\partial t}=\frac {1}{r^2}\frac {\partial
}{\partial r}r^2D(p,r,t) \frac {\partial f}{\partial r} -w\frac
{\partial f}{\partial r}+\frac {\partial f}{\partial p} \frac
{p}{3r^2}\frac {\partial r^2w}{\partial r}
\]
\
\begin{equation}
+\frac {\eta \delta (p-p_{inj})}{4\pi p^2_{inj}m}\rho
(R-0,t)(u(R-0,t)-\dot{R})\delta (r-R(t)).
\end{equation}
Here $\rho$ is the gas density, $u$ is the gas velocity, $P_g$ is
the gas pressure, $P_c=4\pi \int p^2dpvpf/3$ is the cosmic ray
pressure, $w(r,t)$ is the advective velocity of cosmic rays,
$\gamma _g$ is the adiabatic index of the gas, and $D(r,t,p)$ is
the cosmic ray diffusion coefficient. It is assumed that the
diffusive streaming of cosmic rays results in the generation of
magnetohydrodynamic waves and provides the Bohm diffusion
coefficient $D_{B}=\mathit{vpc}/(3\mathit{ZeB})$ for accelerating
particles of charge $\mathit{Ze}$ and velocity $v$ ($c$ is the
speed of light).

The essential feature of our calculations is the inclusion of the
Alfvenic drift effect for particle transport. The Alfven velocity
  $V_{A}=B/\sqrt{4\pi \rho }$ is not negligible in comparison to the
gas velocity {\textmd{$u$} downstream of the shock if the magnetic
field is significantly amplified as indicated by the observations
of the synchrotron X{}-ray radiation. The effective scattering of
energetic particles near the shock is provided by Alfven waves
generated by a resonant cosmic ray streaming instability. Unstable
Alfven waves propagate in the direction opposite to the cosmic ray
gradient. We also assume that the growth of waves is balanced by a
non-linear damping that is strong enough for high wave amplitudes.
The energy which cosmic rays lose due to the wave generation
eventually goes into the gas heating (see the last term in Eq.
(3)). The cosmic ray gradient is negative upstream of the shock
and the waves generated propagate in the positive direction. The
waves transmitted downstream of the shock are damped and
regenerated by the cosmic ray gradient in this region. The cosmic
ray gradient is positive here because of the adiabatic losses of
cosmic ray particles (see numerical calculations of
\cite{Zir08a}). As a result the generated Alfven waves propagate
in the negative direction downstream of the shock.

The situation is somewhat different for the highest energy
particles. An electric current of these particles amplifies the
magnetic field upstream of the shock via a non-resonant streaming
instability \citet{Bell04}. The generated random magnetic fields
are not Alfven waves but almost purely growing magnetic
disturbances. This amplified field plays a role of the regular
field for low-energy particles. They produce Alfven waves that
propagate along the tangled magnetic field lines. The amplified
magnetic field is almost isotropic even downstream of the shock in
spite of compression at the shock front. This is because the shock
front is corrugated by density fluctuations generated by the
streaming instability upstream of the shock \citep{Zir08b}. So the
transport velocity for highest energy particles is close to the
plasma velocity $u$ in contrast to low-energy particles. For the
sake of simplicity we neglect this below and use the Alfven
transport for all energies.

Taking effects mentioned above into account
  we set the cosmic ray advection velocity
equal to $w=u-V_{A}/\sqrt{3}$ downstream of the shock. Because of
this the accelerating particles ``feel'' a smaller compression
ratio and acquire a softer energy spectrum compared to the usual
assumption $w=u$. We set $w=u+V_{A}/\sqrt{3}$ upstream of the
shock to account for the Alfvenic drift effect there. The damping
of Alfven waves created by the cosmic ray streaming instability
results in a very important effect of gas heating upstream of the
shock described by the last term in Eq. (3). This effect limits
the total shock compression ratio.

The spatial dependence of amplified magnetic field is taken in the
form $B(r)=\sqrt{4\pi \rho _0}\frac {\dot{R}\rho }{M_A\rho _0}$
where $\rho _0$ is the gas density of the circumstellar medium and
$M_A$ is some constant. We employ results of \citet{Voelk05} in
the analysis of X-ray radiation from young SNRs and assume that
magnetic energy density $B^{2}/8\mathrm{{\pi}}$ downstream of the
shock is $3.5$ \% of the ram pressure
$\mathit{{\rho}u}_{\mathrm{sh}}^{2}$ that determines the constant
$M_A=23$. It is worth noting that this relation is in good
agreement with the modelling of cosmic ray streaming instability
in young SNRs \citep{Zir08b}.

The last term in Eq. (4) corresponds to the injection of thermal
protons with momenta $p=p_{inj}$ and mass $m$ at the shock at
$r=R(t)$. The dimensionless parameter $\eta$ determines the
injection efficiency. The injection efficiency of thermal ions in
the process of shock acceleration  $\eta =0.1u_{\mathrm{sh}}/c$ is
taken in accordance with the paper \citet{Zir07}; here
$u_{\mathrm{sh}}(t)$ is the time-varying shock velocity.

The maximum particle momentum $p_{\mathrm{max}}$ reached in a
process of diffusive shock acceleration can be roughly estimated
from the condition $D_{B}(p_{\mathrm{max}})\sim
0.1u_{\mathrm{sh}}R_{\mathrm{sh}}$ with $D_{B}$ calculated for the
upstream magnetic field, which is about $5$ times smaller than the
downstream field. This gives an order of magnitude estimate
$p_{\mathrm{max}}c/Z\sim 20(u_{\mathrm{sh}}/10^3\ \mathrm{km} \
\mathrm{s}^{-1})^{2}R_{\mathrm{sh}}\sqrt{n}$ TeV, where the shock
radius is $R_{\mathrm{sh}}$ pc, and the interstellar gas number
density $n$ cm${}^{-3}$.

It can be shown, see e.g. \citet{Ptus05}, that the transformation
of supernova explosion energy to cosmic rays becomes efficient
from the beginning of the Sedov stage of the shock evolution (i.e.
when the mass of supernova ejecta becomes equal to the mass of
swept-up gas) and continues later on. As a result, the
characteristic knee arises in the overall spectrum of particles
accelerated by the evolving supernova remnant. The position of
knee $p_{\mathrm{knee}}$  can be estimated from the above
equations for $p_{\mathrm{max}}$ where $u_{\mathrm{sh}}$  and
$R_{\mathrm{sh}}$ are determined at the time when the Sedov stage
begins. It gives approximately

\begin{equation}
p_{\mathrm{knee}}c/Z\sim 1\cdot
{10}^{15}E_{51}n^{1/6}M_{\mathrm{ej}}^{-2/3} \mathrm{eV}.\ \
\end{equation}

Here $E_{51}$ is the kinetic energy of the supernova explosion in
units of $10^{51}$ erg and $M_{\mathrm{ej}}$ is the mass of
supernova ejecta measured in solar masses.

If the presupernova had a dense star wind with velocity $u_{w}$
and the mass loss rate $\dot{M}$ before the explosion, the shock
may enter the Sedov stage while propagating through the wind
material with mass density $\rho _{w}=\dot{M}/(4\pi u_{w}r^{2})$.
Eq. (5) should be replaced in this case by the following equation:

\begin{equation}
 p_{\mathrm{knee}}c/Z\sim 8\cdot {10}^{15}E_{51}\sqrt{{\dot{M}}_{-
5}/u_{w,6}}M_{\mathrm{ej}}^{-1}
\mathrm{eV}.
\ \
\end{equation}

We have performed numerical simulations of cosmic ray acceleration
for $4$ types of supernova remnants (they constitute about $90$
percents of all supernovae).

$1.$ Type Ia SNRs with the following parameters: kinetic energy of
explosion $E={10}^{51}\mathrm{erg}$, number density of the
surrounding interstellar gas $n=0.1$ cm${}^{-3}$, and mass of
ejecta $M_{\mathrm{ej}}=1.4M_{\odot }$. Also important for
accurate calculations is the index $k$, which describes the power
law density profile $\rho _{s}\propto r^{-k}$ of the outer part of
the star that freely expands after supernova explosions; $k=7$ for
Type Ia supernova.

$2.$ Type IIP SNRs with parameters $E$ = ${10}^{51 }\mathrm{erg}$,
$n=0.1 $ ${\mathrm{cm}}^{-3}$, $M_{\mathrm{ej}}=8M_{\odot }$, and
$k=12.$

$3.$ Type Ib/c SNRs with $E={10}^{51}$ erg exploding into the low
density bubble with density $n=0.01$ cm${}^{-3}$ formed by a
progenitor star that starts off as an O star, goes through a RSG
(Red Super Giant) phase, and ends its life as a Wolf-Rayet star,
see e.g. \citet{Dwar}. The ejecta mass is
$M_{\mathrm{ej}}=2M_{\odot }$ and $k=7$.

$4.$ Type IIb SNRs with $E=3\cdot {10}^{51}$ erg,
$n=0.01$cm${}^{-3}$ and $M_{\mathrm{ej}}=1M_{\odot }$. Before
entering the rarefied bubble, the blast wave goes through the
dense wind emitted by the progenitor star during its final RSG
stage of evolution. We assume that the mass loss rate by the wind
is $\dot{M}={10}^{-4}M_{\odot }/\mathrm{yr}$ and the outer wind
radius is $5$ pc.

A discussion about properties of SNRs produced by core collapse
supernovae can be found in \citet{Chev05}.

We accept the following relative rates for the four types of
supernovae described above: $0.32,{}0.44,{}0.22$ and $0.02$
respectively. The first three rates are taken from the work of
\citet{Smar09} based on the statistics of supernovae within $28$
Mpc of the Galaxy (they also agree with the Galactic supernova
rates derived by \citet{Leam08}). The statistics for rare Type IIb
events is not sufficiently reliable. These supernovae determine
the cosmic ray intensity at energies above about $3\times10^{17}$
eV and our rate $0.02$ is chosen to fit the data on these
ultra-high energy cosmic rays. According to \citet{Smar09}, the
Type IIb supernovae rate can be as high as $0.04$. Notice also
that the fluctuation effect is very strong for these rare events.

The calculated cosmic ray spectra produced over the lifetime of
each type of supernovae are shown in Figure 1 under the assumption
that only protons are accelerated. Here $Q(p)=4\pi p^{2}F(p)$,
where $F(p)$ is the distribution of all accelerated particles
injected in the interstellar medium over the SNR lifetime. The
total number of accelerated particles is $\int Q( p) dp$. It was
assumed that the acceleration ceased at $t_{c}={10}^{5}$ yr. The
shock velocity at this moment is close to $200$ km/s, and the
maximum energy of protons confined in the supernova remnant is
$\sim 5$ TeV. All particles with higher energies accelerated
earlier have left the remnant. The maximum particle energy at late
stages of shock evolution can be two-three orders of magnitude
smaller if one takes into account the possible damping of
turbulence in the shock precursor due to ion-neutral collisions or
non-linear wave interactions \citep{Ptus03}.

The function $F(p)$ was calculated as the sum of two integrals:
the integral taken at $t_{c}$ over the volume of supernova remnant
$(4\pi )\int _{0}^{R_{\mathrm{sh}}( t_{c}) }f(t_{c},r,p) r^{2}dr$,
and the integral over time of the diffusion flux of accelerated
particles through the boundary of the calculation domain $[4\pi\
r^{2}\int _{0}^{t_{c}}(-D \partial f/\partial r)dt]|_{r_{b}}$. The
source function $Q( p)$ should be multiplied by $\nu
_{\mathrm{sn}}$, where $\nu _{\mathrm{sn}}$ is the supernova rate
per unit volume to obtain the density of cosmic ray sources.
Figure 1 shows that about $1/3$ of supernova explosion kinetic
energy $E$ goes to cosmic rays, which
is in agreement with the empirical model of cosmic ray origin.

\begin{figure}[tb]
\includegraphics[width=7.5cm]{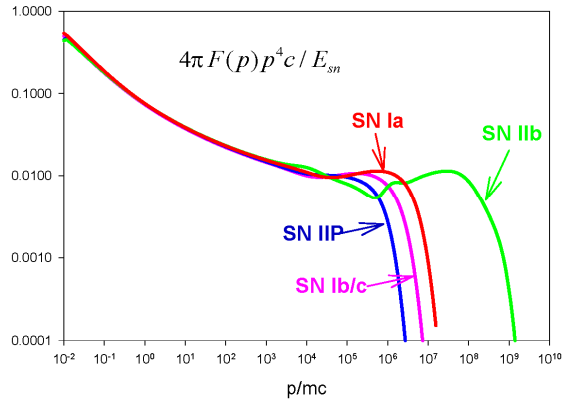}
\caption{Source spectra produced by supernovae Type Ia (solid
line), Type IIP (dash line), Type Ib/c (dotted line), and Type IIb
(dash-dot line) assuming that the accelerated particles are
protons.}
\end{figure}

The spectrum of accelerated energetic ions other than protons has
the same shape if expressed as a function of magnetic rigidity
$Q(p/Z)$ with the appropriate absolute normalization determined by
the injection process at thermal energies.

The relativistic ions released into interstellar space from
numerous supernova remnants, diffuse in galactic magnetic fields,
interact with interstellar gas, and finally escape through the
cosmic ray halo boundaries into intergalactic space, where the
density of cosmic rays is negligible. The main characteristics of
cosmic ray propagation in the Galaxy needed for calculation of the
cosmic ray spectrum in the so called leaky-box approximation is
the escape length $X_{e}$, that is the average matter thickness
traversed by cosmic rays before they exit from the Galaxy, see
e.g. \citet{Strong07}. The cosmic ray intensity obeys the relation
$I\propto\nu _{\mathrm{sn}}Q(X_{e}^{-1}+\sigma/m_{a})^{-1}$, where
$\sigma$ is the nuclear spallation cross section for a given type
of relativistic nuclei moving through the interstellar gas,
$m_{a}$ is the mean interstellar atom mass.

The value of the escape length is determined from the relative
abundance of secondary nuclei (primarily from the Boron-to-Carbon
ratio) in cosmic rays. Based on the paper \citet{Jones01}, we
choose the escape length in the form

\begin{equation}
X_{e}=11.8\left( v/c\right){\left( p/4.9Z \mathrm{GV}\right)
}^{-0.54}\ \ {\mathrm{g}}/{\mathrm{cm}}^{2} \ \
\end{equation}
at $p/Z \geq 4.9$ GV; and $X_{e}\propto v/c$ at $p/Z < 4.9$ GV.
Eq. (7) means that  the resulting spectrum is steeper than the
source spectrum by ${0.54}$ at high enough energies. We use Eq.
(7) at all energies despite the fact that the B/C ratio is not
measured above $100-1000$ GeV.

\begin{figure}
\includegraphics[width=7.5cm]{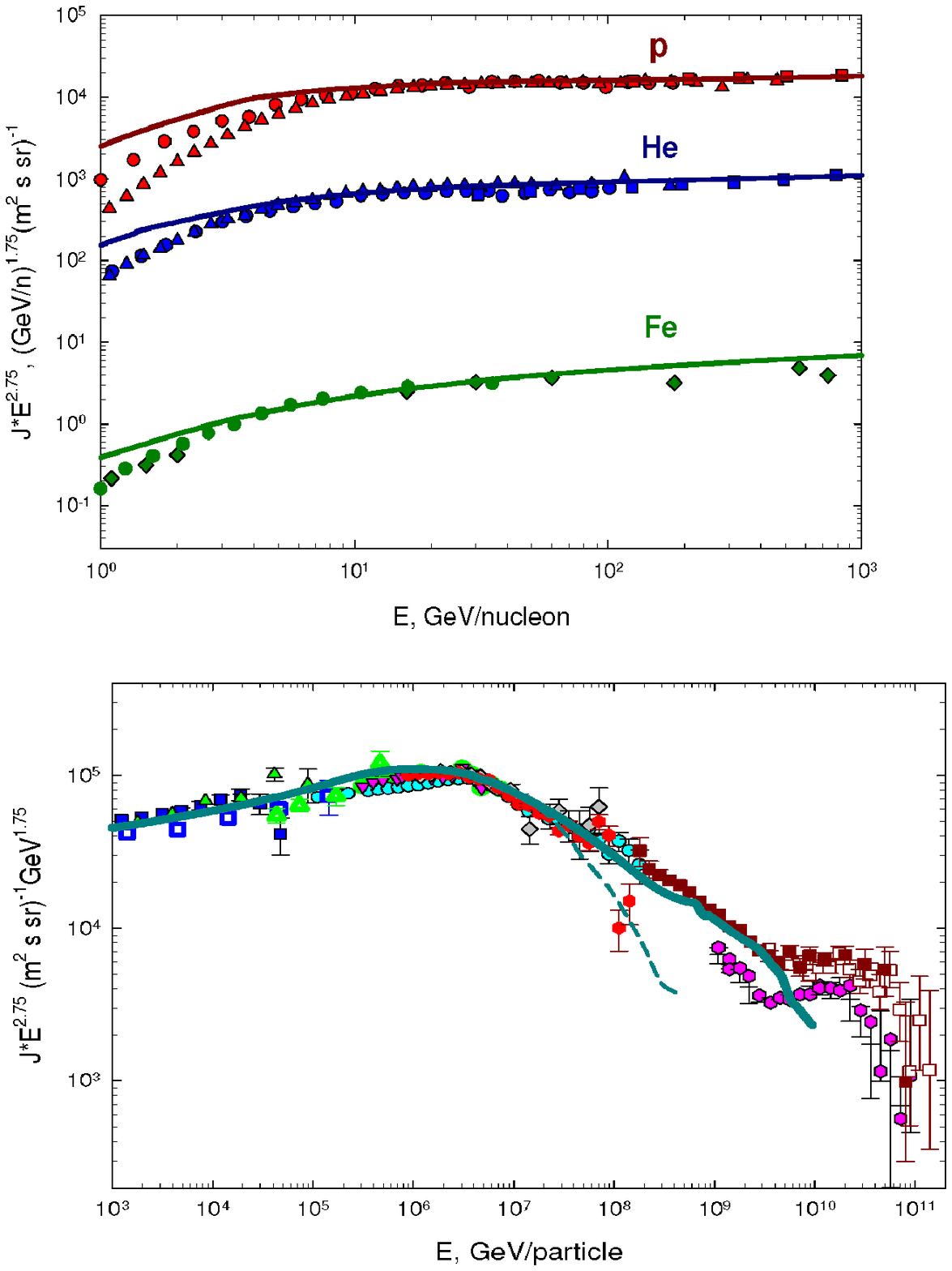}
   \caption{(a) The calculated interstellar (not corrected for solar modulation
at low energies) spectra of protons, Helium, and Iron below energy
${10}^{3}$ GeV/n. Observational data from AMS \citep{Alc00},
ATIC-2 \citep{Pan06}, BESS-TeV \citep{Hai04}, HEAO-3
\citep{Eng90}, and TRACER \citep{Ave08} experiments are shown. (b)
The all particle spectrum above ${10}^{3}$ GeV calculated under
the assumptions that escape length is determined by Eq. (7) at all
energies (solid line) and has a cutoff at $2\cdot{10}^{7}Z$ GeV
(dash line). See \citet{Her09} for references to the observational
data shown by grey symbols.}
 \end{figure}

The source normalization for nuclei from protons to Iron was made
in our calculations from the fit to observed cosmic ray
composition at one reference energy ${10}^{3}$ GeV.

The results of the calculations are shown in Figure 2a for the
interstellar spectra of protons, Helium and Iron nuclei at kinetic
energies per nucleon $1$ GeV/n$<E<{10}^{3}$ GeV/n where the charge
resolution of cosmic ray experiments is high and the escape length
$X_e(E)$ is known. The agreement between our theoretical
predictions and the observations of cosmic-ray energy spectra
supports the validity of our acceleration and propagation models.
(The discordance between the calculated interstellar and the
measured at the Earth spectra at energies below $10$ GeV/n is due
to the solar wind modulation effect.)

The combined spectrum of all protons and ions with particle
energies $E\geq {10}^{3}$ GeV is shown by the solid line in Figure
2b. It was assumed that the charge composition of accelerated
particles was the same in all types of SNRs except that the
highest{}-energy part of the spectrum produced by Type Ib/c
supernovae at $\mathrm{pc}/Z>{10}^{5}$ GeV had no hydrogen. This
reflects the composition of presupernova, the Wolf{}-Rayet star
with a fast H-poor wind. It was assumed that the shock produced by
a supernova explosion first propagates through the $5$ solar
masses of the Wolf{}-Rayet star material uniformly spread in the
central part of the wind bubble and propagates through the
material with normal composition after that. It may explain why
the protons do not dominate in the cosmic{}-ray composition at the
knee \citep{Ant05,Bud07}. In fact the proton{}-to{}-Helium ratio
in cosmic rays is probably decreasing with energy starting at
smaller energies $10^{2}-10^{3}$ GeV/nucleon \citep{Seo07}. We
plan detailed investigation of this point in a separate paper.

The calculated spectra show remarkably good overall fit to
observations up to energies $\sim  5\cdot {10}^{9}$ GeV, where the
transition to extragalactic cosmic rays with the characteristic
GZK suppression above $3\times10^{10}$ GeV likely occurs. To a
good approximation the bending of the observed spectrum at around
the knee energy $3\cdot {10}^{6}$ GeV is reproduced although no
special efforts were made to force the theory to fit the data. The
bending is due to the combined effect of the summation over
different types of SNRs and over different types of accelerated
nuclei.

\begin{figure}[tb]
\includegraphics[width=8.0cm]{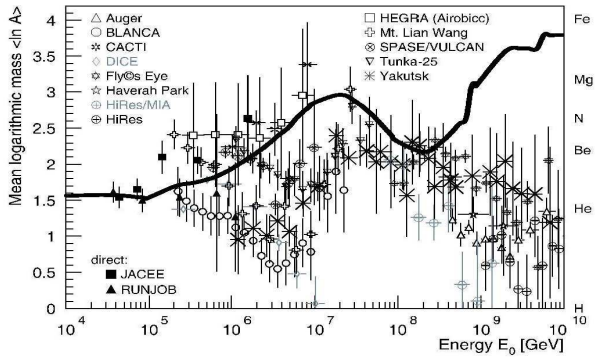}
\caption{Calculated mean logarithmic mass of produced in the
Galaxy cosmic rays (thick solid line) compared to observational
data based on the average depth of shower maximum as presented in
\citet{Her09}.}
\end{figure}

The complicated chemical composition of high energy cosmic rays is
illustrated in Figure 3 where the calculated mean logarithmic
atomic number of cosmic rays $<\ln ( A) >$ is presented. The
increase of $<\ln ( A) >$ at energies from ${10}^{5}$ GeV to
${10}^{7}$ GeV is due to the dependence of the knee position on
charge $p_{\mathrm{knee}}\propto Z$ for each kind of ion
accelerated in Types Ia, IIP, Ib/c SNRs. Type IIb SNRs with normal
composition dominate at rigidities $p/Z>5\cdot {10}^{6}$ GV. They
have a knee at about $p_{\mathrm{knee}}/Z\approx 5\cdot {10}^{7}$
GV and provide progressively heavier composition to the very high
energies. It should be pointed out that the increase of $<ln (A)>$
at $E>3\times10^{8}$ GeV predicted in our calculations is not
supported by the available observations. If confirmed, these
observations may signify the dominant contribution of
extragalactic cosmic rays with light composition at these
energies.

The obtained cosmic ray spectrum shown by the solid line in Figure
2b is very attractive for the explanation of cosmic ray data.
However, the use of the escape length (7) at ultra high energies
is not justified. Experimentally, the value of $X_{e} $ is
determined from the abundance of secondary nuclei in cosmic rays
with good statistics only up to about $100$ GeV/n
\citep{Strong07}. If cosmic ray transport in the Galaxy is
described as diffusion, the diffusion coefficient can be expressed
through the escape length as $D\approx v \mu  H/2X_{e}$ (here $\mu
\approx 0.003$ g/${\mathrm{cm}}^{2}$ is the surface mass density
of Galactic gas disk, $H\approx 4$ kpc is the height of the
Galactic cosmic-ray halo), which gives $D\approx 1.3\cdot
{10}^{28} (p c/Z \mathrm{GeV})^{0.54}$ ${\mathrm{cm}}^{2} $/s. The
diffusion approximation can be used when the diffusion mean free
path $3D/v$ is less than the size of the system $H$, which results
in the condition $pc/Z<2\cdot {10}^{7}$ GeV. At somewhat higher
energies the particles accelerated in the galactic disk fly
straight away from the Galaxy with a flat (source) energy spectrum
and close to hundred percent anisotropy. Certainly this picture
does not represent the reality that may be due to the strong
intermittency of very high energy cosmic rays produced by random
short bursts of not very numerous sources. The dash line in Figure
2b shows the results of calculations made under the assumption
that cosmic ray particles with energies $pc>2\cdot {10}^{7}Z$ GeV
freely escape from the Galaxy without being detected by observer
at the Earth. The predicted spectrum may fit observations only
below about $5\cdot {10}^7$ GeV in this case.

The validity of diffusion approximation extends to higher energies
in the diffusion model with distributed reacceleration on the
interstellar turbulence where $X_{e}\propto (p/Z)^{-1/3}$ at high
rigidities, see \citet{Seo94}. However, this scaling does not
reproduce the observed cosmic ray spectrum for the calculated
source spectrum, see also discussion in Section 3. It is worth
noting that the uncertainty in our knowledge of parameters of the
interstellar turbulence does not allow to decide between two basic
models of cosmic ray propagation: the plain diffusion model and
the diffusion model with distributed reacceleration, see
\citet{Ptusk}.

The physical pattern of cosmic ray propagation is different in the
models with a Galactic wind. The wind model with selfconsistently
calculated cosmic-ray transport coefficients reproduces well the
data on secondary nuclei \citep{Ptus97}. The supersonic wind is
probably terminated by the shock at $\sim 0.5$ Mpc from the
Galactic disk. The confinement of very high energy cosmic rays in
the Galaxy can be more efficient in this model as compared to the
diffusion model with a static flat halo of the size $\sim 4$ kpc
discussed above. In any case, trajectory calculations in galactic
magnetic fields are needed to study cosmic ray propagation when
the diffusion approximation breaks up at ultra high energies. The
detailed consideration of this issue is beyond the scopes of the
present paper.

\section{Discussion and Conclusion}

We have calculated the steady state spectrum of cosmic rays
produced by SNRs in the Galaxy. Our new numerical code
\citep{Zir08a,Zir09} for modelling of particle acceleration by
spherical shocks with the back reaction of cosmic ray pressure on
the shock structure was used in the calculations. The significant
magnetic field amplification in young SNRs inferred from the
observations of their synchrotron X-ray radiations
\citep{Voelk05}, and most probably produced by cosmic ray
streaming instability, was also introduced in the calculations. It
led to the inclusion of the Alfvenic drift in the equation for
particle transport downstream of the shock. Four different types
of SNRs with relative burst rates taken from \citet{Smar09} were
included in the calculations. The escape length Eq.(7) from the
work of \citet{Jones01} was used to describe the propagation of
cosmic rays in the Galaxy in the leaky-box approximation. The
normalization to the observed intensity and chemical composition
of cosmic rays was made at ${10}^{3}$ GeV energy.

The results are illustrated by the solid lines in Figures 2 when
Eq. (7) for $X_{e}$ is used without limitation and by the dash
line when it is limited by the applicability of the diffusion
approximation for cosmic ray propagation in the diffusion model
with a flat static halo. The solid lines reproduce well the entire
cosmic ray spectrum up to $\sim 3\cdot {10}^{9}$ GeV while the
dash line makes it up to less than $\sim 5\cdot {10}^{7}$ GeV.
Further investigations of cosmic ray propagation in galactic
magnetic fields at ultra high energies are needed to refine the
predicted shape of the spectrum produced by the Galactic SNRs at
ultra-high energies. This is important in light of the discussion
about transition from the Galactic to extragalactic component in
the observed cosmic ray spectrum \citep{Berez07,Hillas06}.

Our results can be compared to the earlier work \citet{Ber07},
where the Alfvenic drift effect was not taken into account and
only Type Ia SNRs that represents about $30$\% of all SNRs and can
not efficiently accelerate particles to ultra-high energies were
considered. The absence of Alfvenic drift led to a very flat
cosmic ray source spectrum, which required too strong dependence
of the escape length on rigidity $X_{e}\propto {(p/Z)}^{-0.75}$
inconsistent with the data on secondary nuclei. In fact, this
problem with a too hard predicted source spectrum was the main
motivation for the present work. We included the effect of
Alfvenic drift downstream of the shock and found that the
resulting source spectrum fits the observations. To show the
importance of this effect clearly, we accepted basically the same
set of other parameters and assumptions of the model of shock
acceleration in SNRs as \citet{Ber07}. Also, three additional
types of SNRs were included into the consideration.

The present work demonstrated that supernovae can in principle
produce the source spectrum of galactic cosmic rays required by
the empirical model of cosmic ray origin. More work is needed to
understand how robust are our results.

First of all it is more detailed analysis of MHD effects that
accompany the process of the diffusive shock acceleration in
evolving SNRs. In particular, a further investigation of
generation and transport of MHD turbulence at astrophysical shocks
is necessary. For simplicity we described the properties of MHD
turbulence in our calculations by only one parameter $M_A$.  Its
representative value $M_A=23$ corresponding to the typical
magnitude of amplified magnetic field determined by
\citet{Voelk05} from the observations of non-thermal X-ray
emission from young SNRs gives the particle spectrum that is in
accordance with cosmic-ray observations. However we checked that
this result strongly depends on the assumed value of $M_A$ which
regulates the spectral slope via Alfvenic drift downstream of the
shock. As an example, for $M_A=15$ the spectrum of accelerated
particles is so steep that the escape length $X_{e}\propto
(p/Z)^{-1/3}$ typical for the propagation model with distributed
reacceleration would reproduce the observed cosmic ray spectrum.
The problem is however that the streaming instability of particles
with steep spectrum can hardly generate the strong magnetic
field at large scales that is needded for an acceleration of high-energy
particles.
All this sets one thinking about possible self-consistent
 mechanism that maintains the specific values of $M_A$.

The Alfvenic drift effects downstream of the shock will be
verified in the nearest future via combined observations of young
supernova remnants in Fermi, HESS, MAGIC, VERITAS and some other
experiments. The observations of older remnants will also help to
understand how the maximum energy of accelerated particles depends
on the age of a remnant. Although the maximum energy probably
quickly decreases in the old remnants (see \citet{Ptus03}), we
neglect this effect in our present calculations since it would
require an introduction of an additional rather uncertain
parameter.

A further work is also necessary for determination of the
dependence of injection efficiency $\eta$ on the shock parameters.
The injection efficiency has strong influence on the resulting
particle spectra. It is because the presence of the Alfvenic drift
downstream of the shock steepens the spectrum of accelerated
particles. So, it is more difficult to modify the shock by the
pressure of high energy particles compared to the case without
the Alfvenic drift.

The assumed strong heating in the shock precursor also influences
the resulting particle spectrum. Although the main part of the
cosmic ray energy that goes to the Alfven wave generation is
eventually transformed into the gas heating, the level of Alfvenic
turbulence ($\delta B/B\sim 1$) is high enough to provide the scattering
of cosmic ray particles in the regime close to the Bohm diffusion.
Note that quasi-linear estimates of the Alfven wave
amplitude give $\delta B/B>>1$ for young supernova remnants
\citep{McKenzie82}. In reality Alfven waves generated by the
resonant streaming instability probably stops
when $\delta B/B$ reaches the value close to 1 because of
collisions of the moving adjacent plasma elements similar to results of
\citep{Zir08c}.
The gas heating is then provided by  weak shocks produced in these
collisions.

Our assumption that the composition of accelerated particles is
the same for all types of nuclei (except the very high energy part
of the spectrum produced by Type Ib/c SNRs) and our ignoring the
dispersion of SNR parameters within the same type of supernovae
are probably too simplified to correspond to reality. More
comprehensive analysis taking into account earlier work
\citep{Silb91,Svesh03,Pop07} is needed, although it requires the
introduction of additional not well{}-known astrophysical
parameters.

Another difficult problem is accounting for the fluctuations
resulting from the discrete nature of supernovae in space and
time. Fluctuations of cosmic ray intensity and anisotropy were
addressed in particular by \citet{Erly06,Ptus06}.

\acknowledgments The authors are grateful to referee Michael
Hillas for valuable comments on the manuscript. This work was
supported by the Russian Foundation for Basic Research grant
10-02-00110a. The work was partly fulfilled during VSP visit to the
University of Maryland, USA where it was supported by the NASA
Astronomy and Physics Research and Analysis grant NNX09AC149.


\begin{thebibliography}{}

\bibitem[Aharonian et al.(2007)]{Ahar07} Aharonian, F. et al. 2007, \aap, 464,
253

\bibitem[Alcaraz et al.(2000)]{Alc00} Alcaraz, J. et al. 2000, Phys. Let. B,
490, 27; 494, 93

\bibitem[Antoni et al.(2005)]{Ant05} Antoni, T. et al. (KASCADE
Collab.) 2005, Astropart. Phys., 24, 1

\bibitem[Ave et al.(2008)]{Ave08} Ave, M. et al. 2008, \apj, 678,
262

\bibitem[Bell(2004)]{Bell04} Bell, A.R. 2004,  \mnras, 353, 550

\bibitem[Berezhko \& V\"{o}lk(2007)]{Ber07} Berezhko, E.G. \& V\"{o}lk, H.J.
2007, \apj, 661, L175

\bibitem[Berezinsky et al.(2006)]{Berez07} Berezinsky, V., Gazizov, A. \&
Grigorieva, S.
2006, Phys. Rev. D, 74, 043005

\bibitem[Blumer et al.(2009)]{Her09} Bluemer, J., Engel, R. \& H\"{o}randel,
J.R. 2009, Progress Part. Nuclear Phys., 63, 293

\bibitem[Budnev et al.(2007)]{Bud07} Budnev, N.M. et al. (TUNKA
Collab.) Bull. Russian Acad. Sci.: Physics, 71, 474

\bibitem[Chevalier(2005)]{Chev05} Chevalier, R. 2005, \apj, 619, 839

\bibitem[Dwarkadas(2007)]{Dwar} Dwarkadas, V.V. 2007, \apj, 667,
226

\bibitem[Engelmann et al.(1990)]{Eng90} Engelmann, J. et al. 1990,
\aap, 464, 253

\bibitem[Erlykin \& Wolfendale (2006)]{Erly06} Erlykin, A.D. \& Wolfendale, A.W.
2006,
Astropart. Phys., 25, 183

\bibitem[Haino et al.(2004)]{Hai04} Haino, S. et al. 2004, Phys.
Let. B, 594, 35

\bibitem[Hillas(2006)]{Hillas06} Hillas, A.M. 2006, J. Phys: Conf. Ser., 47, 168

\bibitem[Jones et al.(2001)]{Jones01} Jones, F.C., Lukasiak, A., Ptuskin, V. \&
Webber, W. 2001, \apj, 547, 264

\bibitem[Lagage \& Cesarsky(1983)]{Lag83} Lagage, P.O. \& Cesarsky,
C.J. 1983, \aap, 125, 249

\bibitem[Leaman(2008)]{Leam08} Leaman, J.F. 2008, The Supernova Rate in the
Local Universe, PhD thesis,
 University of California, Berkeley

\bibitem[McKenzie \& V\"olk (1982)]{McKenzie82} McKenzie, J.F., \& V\"olk, H.J.,
1982, A\&A, 116, 191

\bibitem[Panov et al.(2006)]{Pan06} Panov, A.D. et al. 2006, arXiv:0612.377

\bibitem[Popescu(2007)]{Pop07} Popescu, A.S. 2007, arXiv:0704.2718v1

\bibitem[Ptuskin et al.(2006a)]{Ptus06} Ptuskin, V.S., Jones, F.C., Seo, E.S. \&
Sina, R.S. 2006,
Adv. Space Res., 37, 1909

\bibitem[Ptuskin et al.(2006b)]{Ptusk} Ptuskin, V.S., Moskalenko, I.V., Jones,
F.C., Strong, A.W. \& Zirakashvili, V.N. 2006, \apj, 642, 892

\bibitem[Ptuskin et al.(1997)]{Ptus97} Ptuskin, V.S., V\"{o}lk, H.J.,
Zirakashvili, V.N. \&
Breitschwerdt, D. 1997, \aap, 321, 434

\bibitem[Ptuskin \& Zirakashvili(2003)]{Ptus03} Ptuskin, V.S. \& Zirakashvili,
V.N. 2003, \aap, 403, 1

\bibitem[Ptuskin \& Zirakashvili(2005)]{Ptus05} Ptuskin, V.S. \& Zirakashvili,
V.N. 2005, \aap, 429, 755

\bibitem[Seo(2007)]{Seo07} Seo, E.S. 2007, in Cosmic-Rays and High Energy
Universe, ed. T. Shibata,
 N. Sakaki, Universal Academy Press, Inc. - Tokyo, Japan, p. 19

\bibitem[Seo \& Ptuskin(1994)]{Seo94} Seo, E.S. \& Ptuskin, V.S. 1994, \apj,
431, 705

\bibitem[Silberberg et al.(1991)]{Silb91} Silberberg, R., Tsao, C.H., Shapiro,
M.M. \& Biermann, P.L. 1990,
\apj, 363, 265

\bibitem[Smartt et al.(2009)]{Smar09} Smartt, S.J, Eldridge, J.J., Crockett,
R.M. \&
Maund, J.R. 2009, \mnras, 395, 1409

\bibitem[Strong et al.(2007)]{Strong07} Strong, A.W., Moskalenko, I.V. \&
Ptuskin, V.S. 2007, Annu. Rev. Nucl. Part. Sci., 57, 285

\bibitem[Sveshnikova(2003)]{Svesh03} Sveshnikova, L.G. 2003, \aap, 409, 799

\bibitem[V\"{o}lk et al.(2005)]{Voelk05} V\"{o}lk, H.J., Berezhko, E.G. \&
Ksenofontov, L.T. 2005 \aap, 433, 229

\bibitem[Zirakashvili(2007)]{Zir07} Zirakashvili, V.N. 2007, \aap, 466, 1

\bibitem[Zirakashvili \& Ptuskin(2008)]{Zir08b} Zirakashvili, V.N. \& Ptuskin,
V.S. 2008, \apj, 678, 939

\bibitem[Zirakashvili \& Ptuskin(2009a)]{Zir08a} Zirakashvili, V.N. \& Ptuskin,
V.S.
2009,
in High-energy gamma-ray astronomy, eds. F.A. Aharonian et al.,
Melvill, NY, 2009, AIP Conf. Proc. 1085, p. 336

\bibitem[Zirakashvili \& Ptuskin(2009b)]{Zir09} Zirakashvili, V.N. \& Ptuskin,
V.S. 2009, 31st Int. Cosmic Ray Conf.,
 Lodz (Poland), paper HE.1.1

\bibitem[Zirakashvili et al. (2008)]{Zir08c} Zirakashvili, V.N., Ptuskin, V.S.,
\& Volk, H.J.,
2008, \apj, 678, 256

\end{thebibliography}
\end{document}